\documentclass[onecolumn,a4paper,accepted=2023-10-10]{quantumarticle}
\pdfoutput=1
\usepackage[numbers]{natbib}
\usepackage[utf8]{inputenc}
\usepackage[english]{babel}
\usepackage[T1]{fontenc}
\usepackage{amsmath}
\usepackage{hyperref}
\usepackage{amssymb,color}
\usepackage{physics}
\usepackage{booktabs}
\usepackage{listings}
\usepackage{ulem}
\usepackage{authblk}

\definecolor{VPScolor}{rgb}{0.1,0.4,0.9}

\title{Towards Quantum Gravity in the Lab on Quantum Processors }

\author[1]{Illya Shapoval}
\author[2]{Vincent Paul Su}
\author[1]{Wibe de Jong}
\author[1]{Miro Urbanek}
\author[3]{Brian Swingle}

\affil[1]{Lawrence Berkeley National Laboratory, 1 Cyclotron Rd, CA 94720, USA}
\affil[2]{Center for Theoretical Physics and Department of Physics, University of California, Berkeley, CA 94720, U.S.A.}
\affil[3]{Brandeis University, Waltham, MA 02453, USA}

\begin{document}

\maketitle

\abstract{The holographic principle and its realization in the AdS/CFT correspondence led to unexpected connections between general relativity and quantum information. This set the stage for studying aspects of quantum gravity models, which are otherwise difficult to access, in table-top quantum-computational experiments. Recent works have designed a special teleportation protocol that realizes a surprising communication phenomenon most naturally explained by the physics of a traversable wormhole. In this work, we have carried out quantum experiments based on this protocol on state-of-the-art quantum computers. The target quantum processing units (QPUs) included the Quantinuum's trapped-ion System Model \mbox{H1-1} and five IBM superconducting QPUs of various architectures, with public and premium user access. We report the observed teleportation signals from these QPUs with the best one reaching 80\% of theoretical predictions.
In trying to optimize the protocol, we landed on a set of parameters that transfers a classical bit instead of a quantum bit, but the method of transfer still employs quantum scrambling and is an unexpected phenomenon.
We outline the experimental challenges we have faced in the course of implementation, as well as the new theoretical insights into quantum dynamics the work has led to. We also developed QGLab --- an open-source end-to-end software solution that facilitates conducting the wormhole-inspired teleportation experiments on state-of-the-art and emergent generations of QPUs supported by the \textit{Qiskit} and \textit{tket} SDKs. We consider our study and deliverables as an early practical step towards the realization of more complex experiments for the indirect probing of quantum gravity in the lab.}

\section{Introduction}

The twin pillars of quantum field theory and general relativity provide an accurate description of nature across more than 40 orders of magnitude from the tiny constituents of nuclei to the large-scale structure of the universe. However, these two theories of nature have not yet been fused to give a theory of the quantum physics of spacetime and gravity. The search for mathematically consistent models of quantum spacetime and the experimental challenge of determining which model describes nature constitute the problem of quantum gravity.

Experimental access to quantum gravity is challenging at present since it apparently requires the ability to measure miniscule physical effects. Hence, much of the activity in the field to date has focused on formulating mathematically consistent models of quantum gravity. The quantum physics of black holes plays a central role in this search. Indeed, since the theoretical discovery of black hole entropy~\cite{Bekenstein1972,PhysRevD.7.2333,Bekenstein:1974ax,1974Natur.248...30H} (a general result which follows from mild assumptions about the unknown full theory), the community has invested enormous effort into the problem of accounting for the microstates underlying black hole entropy, e.g.~\cite{STROMINGER199699}.

This line of inquiry led to the AdS/CFT correspondence~\cite{Maldacena:1997re,Gubser_1998,https://doi.org/10.48550/arxiv.hep-th/9802150}, which is currently our only candidate for a complete model of quantum gravity. This theory has spurred enormous activity in the field, and has led to unexpected connections between quantum gravity, quantum information, and many-body physics, e.g.~\cite{McGreevy_2010}. However, as presently formulated, AdS/CFT only applies to universes that are qualitatively different from the universe we observe. So the search continues for consistent models of the kind of cosmology we see in the sky.

With experimental access to the quantum properties of the naturally occurring gravitational field still far away, a new idea has emerged: perhaps we can use quantum simulators and quantum computers to indirectly probe quantum gravity in the lab~\cite{Garcia-Alvarez:2016wem,Danshita_2017,Franz:2018cqi,Brown:2019hmk}. Of course, such experiments will not reveal the fine details of the quantum nature of gravity in our universe, but they could shed light on shared structural issues and hard problems arising from the strong coupling dynamics that is likely inherent to any realistic model of quantum gravity. Moreover, quantum many-body systems hosting collective excitations whose most succinct description is in terms of an emergent quantum gravitational field are fascinating and worthy of study in their own right. We also emphasize that these experiments are distinct from studies of analogue gravity, e.g.~\cite{2001gr.qc....11111V}, and from efforts to directly probe the quantum nature of the gravitational attraction that holds us to the Earth, e.g.~\cite{Carney_2019}.

In this paper, we take a practical step towards the goal of quantum gravity in the lab in the context of commercial quantum processors. Following recent proposals in the literature, we have designed and carried out ``wormhole-inspired'' many-body teleportation experiments on IBM and Quantinuum quantum processors and we observe a signal consistent with the predictions in~\cite{Brown:2019hmk,Nezami:2021yaq,Schuster:2021uvg} up to an overall reduction in fidelity. The experimental protocol we consider is inspired by the physics of quantum gravity and, in the right parameter regime, does directly detect holographic wormholes. However, that regime is still inaccesible on current devices, so here we study a simpler situation where the phenomena of interest arise in quantum chaotic systems generally. To be clear, models of quantum gravity are expected to exhibit strongly coupled chaotic dynamics and the protocol itself arose from considerations of quantum gravity, but we do not yet claim to observe an effect unique to quantum gravity.

Our long-term vision is to develop a set of (quantum) computational tools accessible to all interested members of the quantum gravity community. To this end, in addition to our experimental results, we developed a software package built to run our experiments on IBM quantum processors. The package can also be interfaced to Quantinuum systems. In addition to facilities for customized optimization and running the basic experiments, this package includes various methods of error mitigation. Even as the field moves toward fault tolerance, error mitigation will continue to play an important role in practical implementations. We show that, on IBM devices, by optimizing the circuits of interest we can already achieve a signal with 20\% signal fidelity of the ideal result, and we demonstrate that error mitigation can yield another 15\% towards the ideal limit. We also demonstrate that the Quantinuum quantum processor yields a signal with 80\% fidelity of the ideal result without employing any error mitigation techniques. Importantly, the coupling dependence is qualitatively reproduced in both cases, again up to the same overall fidelity reduction.

The concrete data presented in this paper refer to experiments involving 7 and 6 qubits on the IBM and Quantinuum quantum processors, respectively. These were the simplest non-trivial cases we could consider physics-wise given the architectural characteristics and features of the IBM and Quantinuum quantum processors and the run time allocations we had on the platforms. We do observe interesting physics already in this case, and we emphasize that the machinery we have developed can be immediately applied to larger systems or other kinds of interaction topologies as they become available.

The experimental sizes we consider are also comparable to other pioneering experimental studies of quantum scrambling and teleportation, including~\cite{Landsman:2018jpm,Blok:2020may}. In these works, the scrambling unitary was constructed by clever manipulation to enhance the size of local operators. In contrast, our unitary is generated by time evolution of a simple spin chain. One price we pay is that we do not teleport the full quantum state, but rather 
the state with a dephasing channel applied to it.
Another difference is that our protocol does not rely on a probabilistic post-selection on Bell pairs, but rather simply applies a two-sided coupling. Here we emphasize that, unlike those works, we had no control over the target quantum processors and we did not employ pulse-level control in our experiments to achieve the results. For instance, the way we adapted to the error physics in these devices was only to use the provided circuit optimization algorithms and standard error mitigation techniques.

In particular, we faced an interesting experimental challenge: how can we choose the parameters of our model to give the largest possible signal given the architectural and noise constraints. As we discuss in detail below, this experimentally driven process of exploration led to an unexpected result, a regime of the model where we could achieve perfect transmission of 
eigenstates of the Pauli $Z$ operator
in the ideal noiseless setting using the wormhole-inspired protocol. This is still a surprising result, since one does not expect strongly scrambling systems to be able to transmit even classical bits, not to mention qubits. We can apply the protocol to arbitrary quantum states, though the phase information gets lost. We were ultimately able to explain this physics using the same operator-size-based language that underlies the traversable wormhole physics and teleportation-by-size~\cite{Brown:2019hmk,Nezami:2021yaq,Schuster:2021uvg}, but the larger lesson is that the process of developing a practical experimental protocol itself led to new theoretical insights into quantum dynamics. This will surely remain true as we continue to work towards experimental simulations of quantum gravity. Indeed, our findings already raise new questions about potential regimes in AdS/CFT where traversable wormholes can transmit classical but not quantum data. 

The rest of this paper is organized as follows. In Section~\ref{sec:WITP}, we review the holographic teleportation protocol considered in this work. We discuss both the holographic motivation as well as the concrete model we consider, a quantum spin chain introduced by Bertini-Kos-Prosen (BKP) in Ref.~\cite{PhysRevLett.121.264101,Bertini:2018fbz}. We review the ideal case for the BKP model and discuss the particular parameter choice we used in the experiments. In Section~\ref{sec:implementation}, we outline our implementation choices and solutions, including applied circuit optimization and error mitigation techniques, as well as provide motivation for and an overview of the QGLab software tools. Finally, we report the experimental results in Section~\ref{sec:results} and, in Section~\ref{sec:outlook} offer a vision for subsequent research in this direction.

\section{From Quantum Gravity to Qubits}
\label{sec:WITP}

In this section we review the wormhole-inspired teleportation (WIT) protocol that forms the basis for our experimental results. Though its origins come from studying traversable wormholes~\cite{Gao:2016bin,Maldacena_2017} in the AdS/CFT correspondence~\cite{Maldacena:1997re}, recent work \cite{Brown:2019hmk,Nezami:2021yaq,Schuster:2021uvg} has elucidated its quantum mechanical workings by studying a quantity known as \textit{operator size}, which we briefly discuss. A necessary ingredient of WIT is chaotic time evolution. Thus, we also discuss the BKP model~\cite{Bertini:2018wlu,Bertini:2018fbz}, whose time evolution is particularly amenable to implementation on near term quantum devices.

\subsection{Wormhole-Inspired Teleportation}
\label{ssec:protocol}

Let us begin with the existence of a surprising phenomenon. There is an experimental protocol whereby the quantum dynamics of two entangled  systems undergoing chaotic and scrambling time evolution generates an effective channel that teleports quantum information from one system to the other. Whereas chaos tends to be associated with effectively erasing the initial conditions, the transmission of information succeeds precisely when scrambling is maximal. The origins of this protocol stem from the consideration of entangled black holes in AdS/CFT. A negative energy coupling between the pair allows information to teleport from one exterior to the other, a feat that would violate causality without the coupling~\cite{Gao:2016bin,Hartman:2016lgu,Maldacena:2013xja}. Physically, the coupling transforms a pre-existing spatial connection generated by entanglement into a traversable wormhole. As we gain more control over quantum devices, one expects that the CFT side of the duality could in principle be simulated, indirectly probing such wormhole phenomena and other aspects of quantum gravity via the AdS/CFT duality.

A natural question is to ask how far we are from probing such effects on a near term quantum device. Surprisingly, for a handful of qubits, far from a regime where quantum gravitational effects are expected to be relevant, we can directly observe quantum teleportation, albeit with dephasing noise. The lesson here is a surprising one. Rather than the usual quest to discern which quantum mechanical principles survive in a full theory of quantum gravity, a phenomena occurring in quantum gravity has inspired us to search for an analog in quantum mechanics without gravity, leading to rich discoveries about operator size. Put another way, in the pursuit of investigating quantum gravitational phenomena, we learned something new about something we thought we already knew.

\begin{figure}
    \centering
    \includegraphics[width=.8\textwidth]{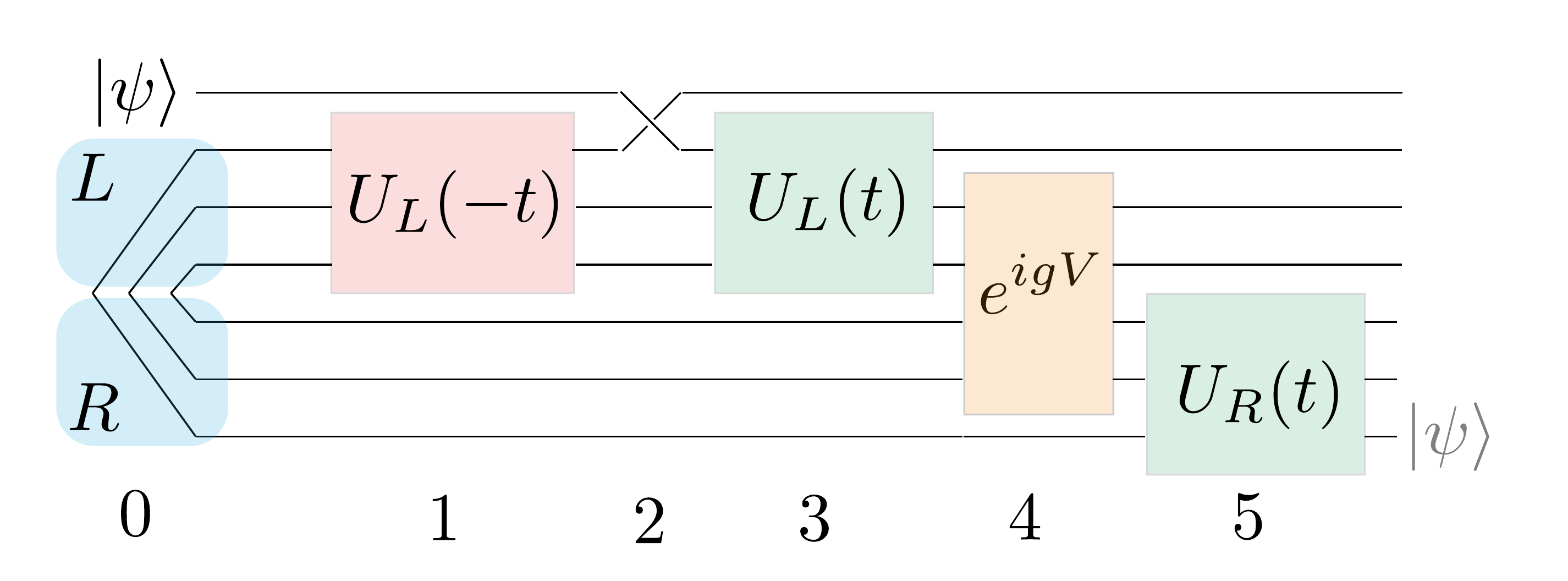}
    \caption{Overview of wormhole-inspired teleportation (WIT) protocol. The key ingredients are preparation of Bell pairs (blue), forwards (green) and backwards (red) time evolution, and a two sided coupling (orange) of the form~\eqref{eq:interaction}. Note that successful teleportation of the state $\ket{\psi}$ can be diagnosed by studying the correlators of time-evolved operators as described in~\cite{Schuster:2021uvg}. For more detailed versions of the circuit for our experimental setup, see Figure~\ref{fig:hl-circuit-8}.}
    \label{fig:protocol}
\end{figure}

We now turn to introducing the WIT protocol that forms the basis for our experimental results in greater detail. The first ingredient, much like the original quantum teleportation protocol, is a number, $n$, of Bell pairs. Half of each Bell pair is sequestered into the ``left'' system consisting of $n$ qubits, leaving a maximally mixed state on the left that is purified by a ``right'' system consisting of the other half of each Bell pair.
Supposing each of the two systems share the same Hamiltonian (up to a transpose in the computational basis), one evolves the left system backwards in time by an amount $t_{0}$. The state $\ket{\psi}$ to be teleported is then swapped with the first qubit and scrambled with the rest of the system by evolving forwards in time again by an amount $t_{0}$. A two-sided coupling of the form $e^{igV}$,
\begin{equation}
  \label{eq:interaction}
   V = \frac{1}{K}\sum_{j}^{K} \sigma^{z}_{j, L}\sigma^{z}_{j, R},
\end{equation}
 acts on the remaining $K=n-1$ entangled partners, sometimes referred to as carrier qubits, on the left and right with $g$ setting the strength of the interaction. By allowing the right system to simply time evolve an amount $t_{0}$, the state $\ket{\psi}$ reappears. Somehow the scrambling dynamics which define the system actually serve to exactly refocus the quantum information to a single register. These steps are summarized in Figure~\ref{fig:protocol}. As noted in \cite{Maldacena:2017axo}, this two sided coupling can be thought of as LOCC between the two sides if the $L$ qubits are measured in the Z basis and the measurement outcomes are used to apply conditional rotations on the $R$ qubits.\footnote{We thank the reviewer for reminding us to include this observation}

This experiment highlights one surprising aspect of quantum dynamics, and it is particularly tantalizing to ask whether this can indeed be used to probe quantum gravitational effects, as proposed in Ref~\cite{Brown:2019hmk}. While we again do not claim to have reached this stage, we outline here two necessary ingredients.
(1) The ability to prepare entangled states on large numbers of qubits. Specifically, access to thermofield double states of the form $\ket{\text{TFD}(\beta)} \equiv \frac{1}{Z_\beta^{1/2}} \sum_n e^{-\beta E_n/2}\ket{n}_L\ket{\bar{n}}_R$ at various temperatures\footnote{In certain systems, preparing the TFD can be recast as a ground state problem~\cite{Cottrell_2019,Maldacena:2018lmt} which  is how Refs~\cite{PhysRevA.104.012427,Maldacena:2019ufo} approach TFD preparation.}. The Bell pair used above is simply the infinite temperature limit $\ket{\text{TFD}(0)}$. 
(2) Implementation of forwards and backwards time evolution for a theory with a gravitational dual, such as $\mathcal{N}=4$ Super Yang-Mills.

Because our code is written in a hardware agnostic way, we hope that it will be immediately useful as future quantum devices with larger quantum volume~\cite{PhysRevA.100.032328} come online. It is of course an open question to find out what other theories have a simple gravitational dual, and the presence of holographic wormholes is one useful diagnostic of a simple dual.

\subsection{Successful Teleportation via Operator Size}
\label{ssec:size}

We now turn to a brief review of teleportation by size, as explained in~\cite{Brown:2019hmk,Schuster:2021uvg}, which provides a microscopic mechanism that enables the WIT protocol to work.

A fundamental notion is the definition of operator size. Consider a generic operator $\mathcal{O}$ on a system of $m$ qubits expressed in the basis of Pauli strings of length $m$.
\begin{equation}
  \label{eq:operator_basis}
  \mathcal{O} = \sum_{P \in \mathcal{P}_{m}} c_{P} P
\end{equation}
We consider both the winding size distribution $q_{\mathcal{O}}(l)$ and the size distribution $Q_{\mathcal{O}}(l)$ of the operator which are defined as
\begin{equation}
  \label{eq:size_distribution}
  q_{\mathcal{O}}(l) = \sum_{|P| = l} c_{P}^{2} \, \quad , \quad \, Q_{\mathcal{O}}(l) = \sum_{|P| = l} |c_{P}|^{2},
\end{equation}
where $|P|$ denotes the number of non-identity terms in the Pauli string $P$. In the case of real coefficients $c_{P}$, as will be the case for our experiment, the two definitions coincide, though the former is expected to be relevant when there is a dual gravitational picture~\cite{Brown:2019hmk}.

Since the state injection on the ``left'' in the WIT protocol can be thought of as an arbitrary perturbation at some time in the past and we would like to compare this to the state emerging in the future on the ``right'', it suffices to study time-evolved operators $\mathcal{O}(t) = e^{+iHt}\mathcal{O} e^{-iHt}$. Assuming that the time evolution is maximally scrambling, one expects that at late times the size distribution of $\mathcal{O}(t)$ has most of its support at size $m$ since the operator would have spread to the entire system.

Finally, it is instructive to understand the effect of the two sided coupling. It is easiest to consider the action again in the Pauli string basis. We start with states of the form $\ket{P}_{LR} = P_R\ket{\phi^+}_{LR}$ with $\ket{\phi^{+}}_{LR}$ the maximally entangled state between the left and right systems as previously described. It turns out that $\ket{P}_{LR}$ is in fact an eigenstate of $V$.

\begin{equation}
  \label{eq:size_coupling}
V\ket{P}_{LR} = (1 - \mathcal{S}_{K,Z}[P])\ket{P}_{LR}
\end{equation}

where $S_{K,Z}[P]$ counts the number of terms in $P$ that do not commute with $\sigma^{z}$ restricted to the $K$ carrier qubits. 
Assuming all Pauli terms are equally likely as would be the case in a maximally scrambled system, in expectation, this gives a measure of the size of $P$ up to some prefactors. Thus, the coupling has the action of applying a size-dependent phase in the Pauli basis. If the size distribution of $O$ is highly concentrated, as would be the case if most of the support is in high-weight Pauli strings, then the state $\mathcal{O}\ket{\phi^+}_{LR}$ would also be an approximate eigenstate of the coupling.

Putting it all together, therefore, when $\mathcal{O}_L(-t)$ is tightly peaked as a result of time evolution on the left side, the two sided coupling applies a global phase that is unwound by the corresponding time evolution on the right side, uncovering the original message sent in on the left side.

A succinct diagnostic for successful quantum teleportation based on the above ingredients was given in~\cite{Schuster:2021uvg}. The two necessary and sufficient conditions are:
\begin{enumerate}
    \item The correlator $\expval{\mathcal{O}_L(-t) e^{igV} \mathcal{O}_R(t)} = re^{i\theta_{\mathcal{O}}}$ has maximal magnitude ($r=1$) for all choices of $\mathcal{O}$.
    \item The phases, $\theta_{\mathcal{O}}$, of each of the correlators are aligned for all operators acting on the message Hilbert space.
\end{enumerate}

When dealing with Bell pairs, the first condition is trivially satisfied. It is the operator phases that are relevant for our experimental teleportation results. As noted, when acting on Bell pairs, these phases are proportional to the size of time evolved operators. In the case of teleporting the state of a single qubit, one needs to examine the set of single qubit Pauli operators. In our model, not all operators will share the same size distribution as a function of time, leading to sub-maximal teleportation. We elaborate on these findings in the next subsection.

\subsection{Theoretical Expectations in the BKP Model}
\label{ssec:BKP}
Since a primary goal of this work is to demonstrate successful teleportation on an available quantum device, we turn to reviewing the self-dual kicked Ising model introduced by Bertini-Kos-Prosen~\cite{PhysRevLett.121.264101,Bertini:2018fbz}. Later in this section, we will also present the analytical and numerical expectations for our quantum experiment. Of course, we are far from probing quantum gravity as outlined in the above requirements. Nonetheless, the BKP model is sufficient for observing successful teleportation in the WIT protocol.
Time evolution in the BKP model is discretized with each time step featuring two alternating components $U = U_{K}U_{I}$ with
\begin{equation}
  \label{eq:time_step}
   U_{K} = \text{exp}{\left(ib\sum_{j}^{n}\sigma_{j}^{x}\right)} \, , \,  U_{I} = \text{exp}{\left(iJ\sum_{j}^{n-1}\sigma_{j}^{z}\sigma_{j+1}^{z} +  i\sum_{j}^{n}h_{j}\sigma_{j}^{z}\right)}
\end{equation}
A nice feature that makes it suitable to implementation on a quantum devices is that both $U_K$ and $U_I$ are individually composed of commuting terms.

Much recent theoretical excitement around this model is due to its maximally chaotic behavior at the ``self-dual'' point, where $|b| = |J| = \pi/4$ and the $h_j$ are initialized as independently distributed random variables~\cite{PhysRevLett.121.264101}.  In the thermodynamic limit, the authors of~\cite{PhysRevLett.121.264101} show that the spectral form factor, the Fourier transform of the two point function of spectral density, in this model matches the expected behavior from random matrix theory, a powerful framework for describing quantum chaotic systems. Another important sense in which it is maximally chaotic is that the entanglement entropy of a subregion grows as fast as possible, namely linearly with time for this one dimensional model~\cite{Bertini:2018fbz}.

Using available quantum simulation packages, we carry out the above WIT protocol using the BKP model. Based on realistic hardware error rates, we considered $n=3$ qubits per side for $T=3$ time steps. We find that the teleportation signal, which we describe in more detail momentarily, is indeed maximized at the self-dual point. A surprising finding was that the particular choice of zero on-site magnetic field $h_j = 0$ also increased the signal, defined as the expected value of $Z$ on the output qubit. In this limit, our teleportation circuit is actually Clifford, allowing for easier analysis of operator size. See Appendix~\ref{app:clifford} for details. From here on, we refer to this set of parameters $\{b = J = \pi/4, h_j = 0, n = 3, T=3\}$ as BKP$_{*}$.

A theoretical calculation using Haar random unitaries, in place of time evolution generated by a Hamiltonian, was carried out in~\cite{Brown:2019hmk}, showing that the effective channel is a depolarizing channel followed by a flip about the Y axis.
\begin{equation}
  \label{eq:depolarizing}
\mathcal{E}_{\text{Haar}}(\rho) : \rho \rightarrow Y\Delta_\lambda(\rho) Y
\end{equation}
where $\Delta_\lambda$ is a depolarizing channel with degradation $1-\lambda$, which depends on the two-sided coupling strength $g$.

Compared to the Haar random case, we find that our choice of time evolution leads to perfect transmission of $Z$ eigenstates in the ideal case for $T=3$ time steps. That is, teleportation of the quantum state with complete dephasing noise. 
Using process tomography (see~\cite{Nielsen00} for a review) to characterize the qubit channel that takes $\rho_{\text{in}}$ to $\rho_{\text{out}}$, we find that
\begin{equation}
\mathcal{E}_{\text{BKP}_{*}}(\rho) : \rho \rightarrow Y \Delta_\lambda(\Phi(\rho)) Y
\end{equation}
where $\Phi$ is a perfectly dephasing channel $\Phi(\rho) : \rho \rightarrow \frac{1}{2}\left(\rho + Z\rho Z\right)$. In other words, eigenstates of $Z$ get flipped with unit fidelity while eigenstates of $X$ and $Y$ lead to a maximally mixed state. The fact that we obtained this channel is directly related to the fact that we tried to maximize the expectation value of $Z$ on the output when the input was a $Z$ eigenstate. This was possible because the circuit had the surprising ability to transmit a classical bit perfectly at the cost of losing coherence. Other choices are also worth exploring and could plausibly lead to other interesting regimes of the dynamics.

Let us emphasize two important points about this channel characterization. The first is that this description is highly sensitive to the experimental parameters we discussed, including model parameters, system size, and number of time steps. If we reduce the number of time steps by 1, the channel would instead dephase in the Y basis rather than the $Z$ (computational) basis. Additionally, in this Clifford regime, the dynamics are periodic and thus, the channel of this circuit does not settle to a ``late time'' limit expected to occur after a scrambling time.
Secondly, we point out that this simplified, though fine-tuned circuit, manages to reproduce the characteristic coupling-dependent depolarization as in the Haar random case. This is a rather non-trivial check that the same physics is at play, even though our circuit does not transmit the phase information. 

\begin{figure}
  \centering
\includegraphics[width=.6\textwidth]{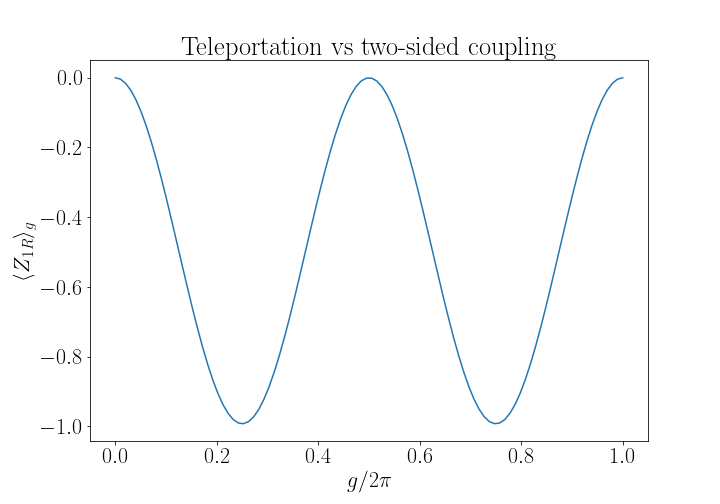}
  \caption{Teleportation fidelity of the state $\ket{\psi} = \ket{0}$ under the WIT protocol (Fig.~\ref{fig:protocol}) with the special set of parameters BKP$_{*}$. The parameter $g$ controls the strength of the two-sided coupling and affects the phases $\theta_{\mathcal{O}}$. Perfect transmission occurs when $\expval{Z}=-1$. When the coupling $g=\frac{\pi}{2}$, phase alignment occurs between $\theta_I = e^{ig}$ and $\theta_Z = e^{i(\pi-g)}$. For the same experimental parameters, both $\expval{Y}$ and $\expval{X}$ are 0. For the same circuit, input states that are eigenstates of the $X,Y$ operators lead to $\rho_{\text{out}} = I$, which fails to teleport any information.}
  \label{fig:analytic_teleportation}
\end{figure}

Interestingly, this case of sub-perfect teleportation can be precisely understood by analyzing the conditions laid out in the previous subsection. For our circuit that teleports one qubit's worth of information, we need to analyze correlators for the single qubit operators $\{I,X,Y,Z\}$. Condition 1 is trivially satisfied, but the phases are not aligned for any choice of $g$. This is because the sizes of the time-evolved operators do not match. We list time-evolved operators in Appendix~\ref{app:op_growth}.
This is analogous to the late time, high temperature regime described in~\cite{Schuster:2021uvg}. However, note that the Clifford dynamics are actually periodic, so there is not a well defined notion of scrambling time, and thus no well-defined notion of late time in terms of the operator growth.

To simplify the necessary measurements for the experiment, we take the input state $\ket{\psi}$ to be the plus eigenstate of the $Z$ operator, $\ket{0}$. The output fidelity can then simply be measured as the expectation value of the $Z$ operator, where perfect fidelity would result in an expectation value of $\expval{Z}_{\ket{\psi}_{\text{out}}} = -1$. We then calculated the expectation value of $\expval{Z}$ as a function of the two sided coupling $g$, finding that indeed it is maximized at $g=\pi/2$, which coincides with phase alignment between $Z$ and the identity operator $I$. See Figure~\ref{fig:analytic_teleportation}. Having these phases aligned is significant because the density matrix for $\ket{\psi} = \ket{0}$ can be expressed as a linear combination of $I$ and $Z$.



That this quantum chaotic circuit transmits classical information is rather unexpected. 
One interpretation of these results is that it suggests that holographic teleportation may not always correspond to perfect quantum teleportation. In the case of holography, we expect this transmission to correspond to travelling through a wormhole. However, it is possible to engineer scenarios where the infalling particle interacts with some other matter behind the horizon, leading to a type of decoherence similar to one as we have observed. For example, if one starts not with the TFD state but a slightly perturbed one, the interior geometry may no longer simply be the vacuum. Carrying out the protocol with these different initial conditions is the so-called ``wormhole tomography'' suggested in~\cite{Brown:2019hmk}.

\section{Implementation}
\label{sec:implementation}

We now turn to the implementation of the WIT on quantum processors. As the global race for scalable quantum processors heats up, the manifold of digital quantum processors fabricated by research labs, startups and technology companies worldwide is rapidly expanding. In this work, we used quantum systems from IBM and Quantinuum to experiment with the WIT. On the one hand, we chose IBM Quantum~\cite{IBMQQ} for its full quantum stack embracing a wide family of superconducting quantum processor units (QPUs), which are available through IBM Cloud\textsuperscript{\textregistered}, and the open-source software development kit for programming and using the QPUs. Furthermore, our access to the premium QPU lineup through the IBM Quantum Network opened additional degrees of freedom. On the other hand, we countered that with a Quantinuum system based on trapped-ion qubits, which provided us with leading-edge error rates and all-to-all topology. In this section, we will introduce the QPUs of our choice and will walk through the main aspects of adapting the protocol to the architectures of our choice.

\subsection{Quantum processors}

\subsubsection{IBM Quantum processors}
\label{ssec:ibm_qpus}

State-of-the-art IBM quantum systems implement an electrically controlled solid-state quantum computer. The implementation is based on fixed-frequency superconducting transmon qubits. The set of native gates, implemented in microwave pulses, consists of single-qubit \textit{I}, \textit{RX}, \textit{RZ} and two-qubit \textit{CX} gates. Select pairs of qubits are coupled by coplanar waveguide bus resonators thus forming a lattice of a certain topology. The \textit{CX} gate can only act on directly connected qubits, imposing the lattice topology to be a circuit-routing constraint in cases when the algorithmic coupling graph is not a subgraph of the QPU lattice.

\begin{figure}[ht]
    \centering
    \begin{minipage}[t]{.45\textwidth}
    \vspace{-20mm}
    \includegraphics[scale=0.25]{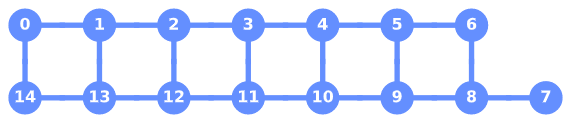}
    \end{minipage}
    \begin{minipage}[t]{.45\textwidth}
    \includegraphics[scale=0.25]{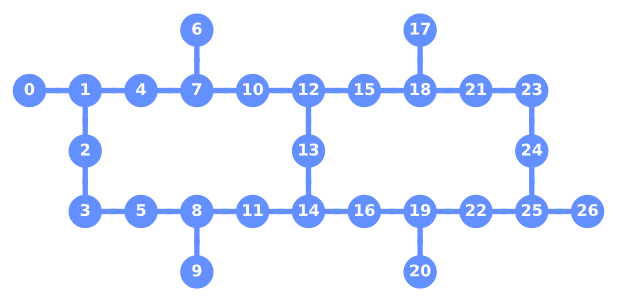}
    \end{minipage}
    \caption{The square-lattice architecture of the Canary-family public-access 15-qubit \textit{ibmq\_16\_melbourne} chip (left) and the heavy-hexagonal lattice of the Falcon-family premium-access 27-qubit \textit{ibmq\_montreal} chip (right) featuring lower-degree vertices.}
    \label{fig:montreal_topology}
\end{figure}

Figure~\ref{fig:montreal_topology} shows two types of lattices we experimented with that are notably different in terms of topology satisfiability for our purposes. The square lattice of \textit{ibmq\_16\_melbourne} – one of the early-generation IBM Quantum Canary processors – matches perfectly the coupling requirements of the WIT protocol. In general, a perfectly matching topology allows to avoid the qubit routing overhead and so, with all other conditions being equal, yields better experimental results. However, \textit{ibmq\_16\_melbourne} suffered from relatively high error rates that were difficult to mitigate, leading to mediocre results in our experiments. On the other hand, the heavy-hexagonal lattice\footnote{The heavy-hex lattice is the fourth and the latest generation of topology for IBM QPUs, which has been made the basis for all in-production processors with 16 or more qubits since August 2021.} of the Falcon-family \textit{ibmq\_montreal} processor led to superior signal yield for our protocol. In this case, the unavoidable routing overhead in the number of gates caused by the lattice topology mismatch was entirely offset by the benefits of lower-sparsity heavy-hex topology, the most crucial of which were increased zero-frequency collision yield~\cite{Hertzberg_2021} and improved overall gate fidelity~\cite{Takita_2017}. These improvements, as well as others in gate design, qubit readout, and control software, allowed \textit{ibmq\_montreal} to reach the IBM-stated quantum volume of 128 --- the highest quantum volume across all accessible IBM devices to date\footnote{In April of 2022, IBM announced achieving a quantum volume of 256 on the Falcon-family \textit{ibmq\_prague}. However, the processor is not yet publicly accessible at the time of this writing.}. These considerations made \textit{ibmq\_montreal}, as well as three other heavy-hexagonal processors of lower quantum volume --- \textit{ibmq\_paris}, \textit{ibmq\_manhattan}, \textit{ibmq\_toronto}\footnote{All experiments on IBM processors were run over the period of several months, until September 2021, spanning multiple generations of processors. As of this writing, \textit{ibmq\_16\_melbourne}, \textit{ibmq\_paris}, and \textit{ibmq\_manhattan} are decommissioned by IBM.} --- our target devices for this study.

\subsubsection{Quantinuum processor}
Quantinuum's universal quantum computers use moveable ions as qubits within a trapped-ion quantum charge-coupled device~\cite{Pino2021}. All experiments were done on Quantinuum's 10-qubit linear-architecture System Model H1-1, powered by Honeywell. Its gate set consists of single-qubit \textit{RX}, \textit{RY}, \textit{RZ} and two-qubit \textit{ZZ} gates. \textit{CX} gate can be readily implemented through a combination of one-qubit operations combined with one \textit{ZZ}. In addition, the system has a highly accurate reset gate operation available to its users. The trapped ion has an all-to-all qubit connectivity, with ions being shuttled to enable direct connections with all qubits. The H1-1 has demonstrated average fidelities of 99.99\% for single-qubit operations, and 99.7\% for two-qubit gate operations and state preparation and measurement, claiming a quantum volume of 1024~\cite{H1QVFidelity}.

\subsection{Circuit compilation and optimization}
\label{ssec:circuit-compilation}
In section~\ref{ssec:protocol}, we formulated the protocol as a hardware-agnostic abstraction. To execute the protocol on quantum hardware, the high-level circuit representation must be compiled to target device to account for the physical constraints and properties of its architecture. The bare minimum of such compilation includes three phases: decomposition of high-level gates to basis gates, initial mapping of logical qubits to physical ones, and routing the circuit to account for the qubits' coupling constraints, if any. Optionally, the main phases can be interleaved with various optimizations making compiled circuit shallower. Finally, the mapping and routing phases can be performed under the requirement of minimizing the amount of noise the compiled circuit would accrue from gate and measurement errors. Circuit optimization and the noise-aware mode of mapping and routing, despite being optional, are often determinant for the quality or even feasibility of a quantum experiment on the NISQ QPUs. The extent of optimization and noise minimization turned out to be critical for the success of our experiments on the IBM systems.

\begin{figure}[ht]
    \centering
    \includegraphics[scale=0.4]{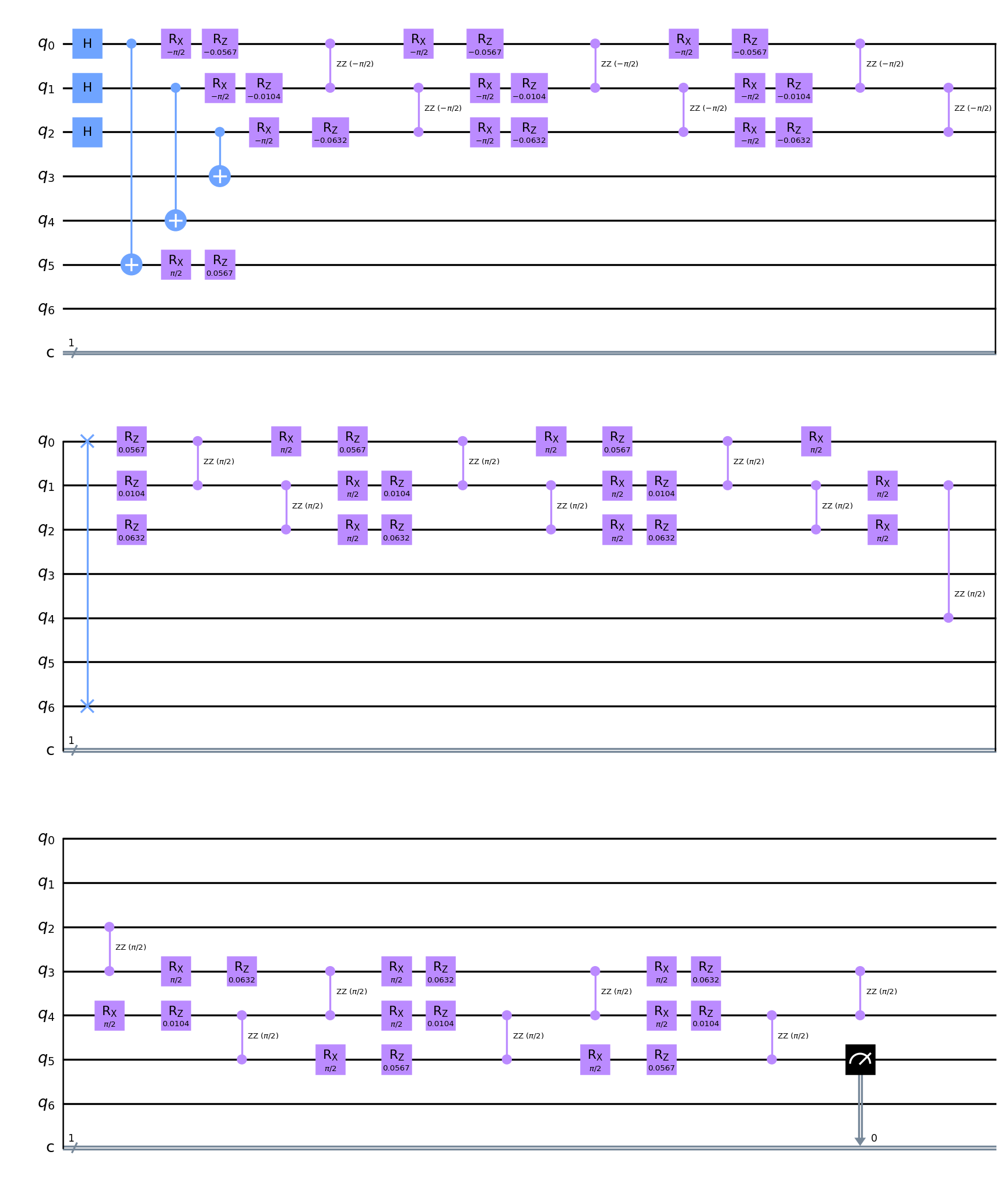}
    \caption{High-level WIT quantum circuit at $g=\pi/2$ with 3 qubits per side for $+Z$ initial state and final $Z$ measurement. Message is inserted with the \textit{SWAP} gate (blue line connecting two crosses). [Drawn with IBM's Qiskit SDK]}
    \label{fig:hl-circuit-8}
\end{figure}

State-of-the-art quantum compilers, such as those provided by \textit{Qiskit} and \textit{tket} SDKs, can automate the above-mentioned circuit translations to a certain degree. Default turnkey compilation workflows yield excellent results for simple input circuits. In general, however, finding an optimal compilation solution for a circuit of arbitrary type and complexity on the one hand, and faulty and constrained hardware on the other, is non-trivial and, in practice, heuristic. In this light, in addition to the main goal of demonstrating the WIT protocol on the cutting-edge hardware, we also aimed to evaluate the current capacity of available software to compile optimally --- either out-of-the-box, or with minimal tweaks --- the circuits of nature and size similar to the one of our interest (see Figure~\ref{fig:hl-circuit-8}), and to create a reference point for related circuit extensions that might be of further interest for quantum gravity studies.

\subsubsection{Compiling for IBMQ Montreal}

Table~\ref{tab:compile-to-IBM} summarizes the compilation outputs obtained with \textit{Qiskit} and \textit{tket} over a set of configurations on \textit{ibmq\_montreal}, the processor that we found to be the most capable in the IBM lineup for our purposes. The researched phase space spanned default and moderately tuned workflows, as well as more intrusive workflow re-configurations.

\begin{table}[ht]
    \centering
    \begin{tabular}{l c c c c c}
    \toprule
         &  \multicolumn{5}{c}{\textbf{Compilation output}}\\
         \cmidrule(r){2-6}
         & \textit{CX} & \textit{SX} & \textit{RZ} & Depth & Size \\
         \cmidrule(r){2-6}
         A: \textbf{Qiskit}: default passes & 73--92 & 64--80 & 83--96 & 115--147 & 227--265 \\
         B: \textbf{tket}: default passes & 49 & 83 & 103 & 131 & 237 \\
         C: \textbf{Qiskit}: reconfigured case A & 39 & 59--78 & 69--96 & 72--90 & 171--214 \\
         D: \textbf{tket}: custom pass sequence & 38 & 82 & 123 & 123 & 244 \\
         E: \textbf{Hybrid}: tket \& Qiskit & 34 & 64--69 & 81--86 & 84--90 & 183--190 \\ 
    \bottomrule
    \end{tabular}
    \caption{Gate counts of the WIT circuit (see Figure~\ref{fig:hl-circuit-8}) compiled for \textit{ibmq\_montreal}. Optimization levels in all stages of \textit{tket} (v0.11.0) and \textit{Qiskit} (v0.26.2) are maximal. The outcomes are ordered by \textit{CX} counts --- the major factor for experiment fidelity. Ranges in counts reflect variability of output over 14 compilation trials, with the variability arising from stochastic techniques of the \textit{Qiskit}'s compiler. Chaining \textit{tket} and \textit{Qiskit} lead to the best compilation solution.}
    \label{tab:compile-to-IBM}
\end{table}

The results are ordered by \textit{CX} gate counts. The worst compilation output, solution A, was generated by \textit{Qiskit} with its default pass manager in default configuration. No teleportation yield --- neither raw nor error-mitigated --- was measured on \textit{ibmq\_montreal} in this case due to the amount of noise accrued by the compiled circuit. In solution B, the improvements in gate counts brought by \textit{tket}’s default sequence of passes were still not sufficient to overcome the noise. It is with solution C that we were able to extract a weak teleportation signal with error-mitigation enabled. The change became possible with (1) reconfiguration of the pass manager used in case A to use SABRE~\cite{Li:2019} --- the iterative swap-based bidirectional heuristic algorithm for routing of the input circuit, and (2) selection of a solution with minimal \textit{CX} count across a thousand of identically configured compilations. We also implemented a custom sequence of passes for \textit{tket} eliminating 11 \textit{CX} gates as compared to the \textit{tket}’s default sequence of passes (solution D). Finally, the best solution (E) (see Figure~\ref{fig:compiled-circuit-8} of Appendix~\ref{app:compiled-circuit}) was generated, somewhat unconventionally, chaining both compilers: first, applying \textit{tket} in configuration D, followed by \textit{Qiskit} in configuration A.

\subsubsection{Compiling for Quantinuum System Model H1-1}

The high-level WIT circuits (see Figure~\ref{fig:hl-circuit-8}) generated by QGLab (see Section~\ref{ssec:qglab_soft}) were compiled for H1-1 manually. Firstly, we adapted the circuits to the processor's gate set. Secondly, the fast mid-circuit reset gate featured by the processor allowed to reduce the number of involved qubits from seven to six by eliminating the swap gate used for message insertion on IBM QPUs. Also, the all-to-all topology of the processor allowed to avoid qubit routing. Finally, we did not apply any gate-level optimizations to the circuit on this processor. The compiled circuits thus contained 43 \textit{CX} gates and preserved the original square-lattice structure of the protocol. In Section~\ref{sec:results}, we will see whether the advantage of the best solution for the IBM Montreal QPU in terms of the number of \textit{CX} gates (see Table~\ref{tab:compile-to-IBM}) transformed into any advantage in experimental signal yield on the IBM processor.

\subsection{Error-mitigation techniques employed on the IBM QPUs}
\label{ssec:error_mitigation}

Current quantum computers are sensitive to various errors. These errors can be broadly separated into coherent and incoherent errors. Coherent errors are typically caused by slight miscalibrations of gate parameters and lead to under-rotations or over-rotations of gate actions on qubits. They can be characterized by unitary matrices and are typically repeated in series of experiments. On the other hand, incoherent errors typically originate from interactions between the system and the environment. They are stochastic and transform a pure quantum state into a mixed state.

Another source of error are state preparation and readout errors. We assume that all qubits in the system are prepared in the perfect zero state. This is not always the case as there may exist a small population of excited states. Readout errors are caused by a measurement apparatus. In superconducting qubits, there is a balance between the duration of the measurement and its uncertainty. Long measurement lead to decay of qubits, so the measurement duration has to be properly tuned. There is also a certain chance of measuring the incorrect qubit outcome due to small overlap between probability distributions of the measured physical quantities.

In this study, we employed the following techniques to mitigate errors in experiments on the IBM QPUs:
\begin{itemize}
\item \emph{Heralding}. A measurement of all qubits is performed at the beginning of a circuit. If the outcome is not the zero state, the final results of the circuit execution is discarded. It can be alternatively implemented as active qubit reset. Heralding mitigates errors in state preparation.
\item \emph{Readout-error correction}. Readout errors are caused by statistical uncertainty and long duration of measurement. To correct them, we first measure a readout response matrix, i.e., we prepare each computational state and sample the outcome statistics. Then we use the response matrix to correct the measured statistics for the target circuit~\cite{Kandala:2017, Nachman:2020}.
\item \emph{Randomized compiling}. This technique transforms coherent errors into incoherent errors that can be mitigated more efficiently. We replace each \textit{CX} gate by a pair of single-qubit gates, a \textit{CX} gate, and another pair of single-qubit gates acting on the respective qubits. The single-qubit gates are picked from \{\textit{I, X, Y, Z}\}, where $I$ is the identity gate, and \textit{X}, \textit{Y}, and \textit{Z} are the Pauli gates. They are chosen randomly, but in a way that the total action of all five gates is again a \textit{CX} gate. The ideal circuit output therefore does not change, but each individual \textit{CX} is now applied in a randomized frame. By executing many randomized instances of such circuits, we effectively convert coherent errors in the \textit{CX} gates into incoherent errors~\cite{Wallman:2016}.
\item \emph{Mitigation with estimation circuits}. We replace all \textit{CX} gates in the circuit with identities to create an estimation circuit from the target circuit. We then use the estimation circuit to measure the global noise rate and use the noise rate to correct an output of the target circuit~\cite{PhysRevLett.127.270502}.
\item \emph{Zero-noise extrapolation}. With this method we vary the amount of noise by replacing each \textit{CX} gate with three and five \textit{CX} gates. We execute the modified circuits to measure expectation values with increased circuit noise. Then we extrapolate to the zero-noise limit to obtain a corrected expectation value~\cite{Li:2017, Temme:2017, Kandala:2019}.
\end{itemize}
Heralding, readout-error correction, and randomized compiling were applied to each executed circuit. We performed mitigation with estimation circuits on each circuit required for zero-noise extrapolation. The zero-noise extrapolation was used to obtain the final mitigated values. Detailed discussion of the mitigation protocol can be found in Ref.~\cite{PhysRevLett.127.270502}.

\subsection{QGLab: automating holographic teleportation experiments}
\label{ssec:qglab_soft}

The nature and architecture of state-of-the-art quantum processors and services entail several peculiarities in assembling and postprocessing a quantum computation as compared to its classical counterpart. In particular, the assembling stage is more ambiguous, hardware-dependent, and time-bound. For example, a compilation solution for a given processor may degrade over time in terms of associated computation yield due to the inherent drift of the hardware. For circuits of sufficient size and noise accrual power, the effect may undermine experimental results or even wipe them out in cases when the measured quantities are theoretically marginal. This prompts just-in-time revaluation of compilation solutions for each experiment session. The problem of a compilation solution going obsolete can, however, extend further due to the QPU contention among the ever-growing number of users of cloud-based quantum services, which results in the execution time being delayed for hours, days, or, sometimes, weeks. In addition, a user of cloud-based quantum service faces relatively frequent QPU firmware upgrades, the retirement of old and releases of new QPU generations. All aforementioned factors necessitate the ability to switch between processors having, in general, incompatible topology or basis gates with minimal effort --- preferably, in an automatized way --- while maximizing the expected computation outcome. The dimensions of scalability and extensibility of experiments, hardware-specific job partitioning call for more automation, which, if implemented, would make the physical experiments more streamlined.

We developed QGLab~\cite{QGLAB:2021} --- an open-source end-to-end software solution that facilitates conducting the holographic teleportation experiments on state-of-the-art and emergent generations of QPUs supported by the \textit{Qiskit} and \textit{tket} SDKs. The code bundles all stages of an experiment as a single configurable workflow. This allows for faster development and experimentation cycles under the above-outlined characteristic traits of the quantum computing environment. The workflow comprises the following stages:
\begin{enumerate}
    \item Configure compiler, optimization methods, and output resolution.
    \item Interactively select a QPU based on live occupancy.
    \item Compile experiment’s circuits (optionally, with search for supersolutions)
    \item Prepare and submit batch jobs to the cloud (optionally, asynchronously).
    \item Fetch and post-process the results.
\end{enumerate}
The listing in Appendix~\ref{app:QGLabLog} shows a typical command output of the workflow.

The compilation stage optionally invokes an additional mechanism for maximizing the yield of an experiment on target QPU, extending conventional quantum compilation steps. This was motivated by the observation that both graph-based and noise-adaptive solutions derived by \textit{Qiskit} and \textit{tket} compilers in the default setting often turned out to be suboptimal. Instead of letting the compilers produce a single solution, the mechanism requests a set of solutions satisfying a chosen placement strategy and ranks them by the deviation of the experiment’s outcome, obtained in a noisy simulation, from the noiseless prediction. The mechanism allowed us to systematically improve the results on the type and size of circuits of our interest. Interestingly, being guided by this feature, we often observed solutions of somewhat higher routing overhead producing better results in the actual quantum experiments.

The code also provides the following benefits:
\begin{enumerate}
    \item Ease of installation.
    \item Ease of switching between QPUs supported by \textit{Qiskit} and \textit{tket}.
    \item Ease of experiment resolution scaling (based on automatic jobs’ batching).
    \item Ease of scaling to more qubits and time steps.
    \item Automatic readout error mitigation.
    \item Automatic reproducibility analysis.
    \item Ease of plugging new types of experiments.
\end{enumerate}

Our vision is that the development of open-source specialized software for embedding, driving, scaling and replicating experiments related to quantum gravity on NISQ QPUs may benefit the community accelerating further progress in this field. We hope QGLab is one of the first steps toward this goal.

\section{Results}
\label{sec:results}

\begin{figure}[h]
    \centering
    \includegraphics[scale=0.4]{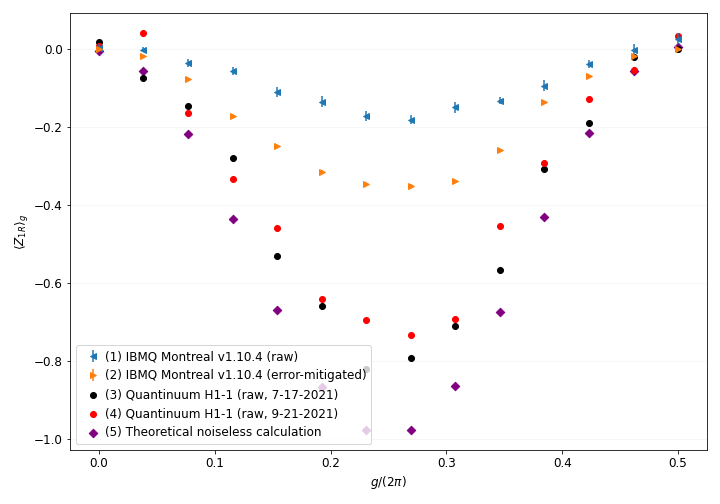}
    \caption{Experimental WIT as a function of two-sided coupling $g$ measured on the IBM \textit{ibmq\_montreal} v1.10.4 and the Quantinuum System Model H1-1. We compare the results to the theoretical, noiseless calculation. (1) Each data point and error bar are the mean and the standard deviation, respectively, of a set of 10 $\expval{Z}$ measured in 10 independent experiments, without error mitigation. The mean from each experiment is calculated over 8192 shots. (2) The error-mitigated data is processed as per the protocol described in Section~\ref{ssec:error_mitigation}. The error bars are the standard error of processed data. (3,4) The two curves correspond to two series of experiments without error mitigation run on 7-17-2021 (3) and 9-21-2021 (4). Each data point is the mean over 1000 shots.}
    \label{fig:results}
\end{figure}

In Figure~\ref{fig:results}, we present and compare our best results obtained from experiments on the IBM and Quantinuum QPUs.

We ran the WIT experiments on five IBM processors (see Section~\ref{ssec:ibm_qpus}) representing square-lattice and heavy-hex topologies, and three processor families: Canary, Hummingbird, and Falcon. The QGLab automation tools introduced in~\ref{ssec:qglab_soft} facilitated the exploration of the compilation solution space and selection of the best solutions on each of the processors. The teleportation signal produced by the \textit{ibmq\_montreal} --- the strongest and most stable signal across all the target IBM processors, both in raw and error-mitigated yields --- is shown in Figure~\ref{fig:results}. Application of the error mitigation techniques outlined in subsection~\ref{ssec:error_mitigation} reproducibly improved the raw outcome by about 93.8\% at the peak of the signal, with dominant corrections coming from randomized compiling and mitigation with estimation circuits. However, the error-mitigated signal on this processor was still about 65\% lower than the theoretical prediction at the peak.

In contrast to experiments on the IBM QPUs, the ones on H1-1 had fewer runs and no error mitigation due to our limited run time allocation on the QPU. Two sets of independent experiments, with 1000 shots each, were run for every two-sided coupling $g$. The result exceeded four-fold the corresponding raw outcome obtained on \textit{ibmq\_montreal} reaching about 80\% of the theoretical prediction.

Higher-fidelity gates and measurement, and longer coherence times of H1-1 were likely the dominant drivers of significant improvement in the raw teleportation signal yield. Note that the experiments on H1-1 involved six more \textit{CX} gates as compared to experiments on the IBM Montreal. Subsequently, we were able to reduce the number of \textit{CX} gates to 20 when compiling for the H1-1 with QGLab/tket while factoring in the processor's all-to-all topology. Thus, we expect fully matching the theoretical prediction in the WIT parameter regime considered in this work to be relatively straightforward on the H1-1, even without error mitigation.

\section{Outlook}
\label{sec:outlook}

These results are an early step towards investigating quantum gravity in the lab, but based on our results and experience with these machines, we conclude that we have likely reached a point where it becomes possible to obtain both qualitatively and quantitatively significant results from even more advanced experiments in this direction. For example, we believe that the H1-1 generation at its current quantum volume of 1024 may already allow to scale up the experiments outlined in this study in the number of qubits and time steps. 

One upcoming milestone we envision on this journey is performing the experiment with access to finite temperature TFD states, dual to two-sided black holes~\cite{Maldacena:2013xja}. For recent work on preparing TFD states in the context of near-term quantum devices, see~\cite{doi:10.1073/pnas.2006337117,PhysRevLett.123.220502,PhysRevA.104.012427}. For example, the Quantinuum System Model H1-2 --- the second generation H1 processor that has recently demonstrated a quantum volume of 2048 --- may make extension of the WIT experiments with finite temperature TFD initial states feasible.

For the purposes of probing quantum gravity, we expect that a larger number of qubits will be important as well, though it may also be interesting to use the teleportation protocol as a benchmark of coherent evolution on future devices. Some additional interesting questions on the gravitational side include searching for theories with simple holographic duals, testing the sensitivity of the protocol to initial conditions, looking for bulk avatars of boundary noise, and probing behind-the-horizon physics.

Finally, it is interesting to better understand the information transmission phenomenon we observed. For example, one could use our teleportation with dephasing noise as a means to do perfect bit transfer. Can we arrange a similar situation in AdS/CFT and are there any advantages to transmitting classical data using quantum scrambling? Also, can we understand how the coherence in the $Z$-basis is hidden in the overall many-body state?

\subsection*{Acknowledgments}
This research used resources of the Oak Ridge Leadership Computing Facility, which is a DOE Office of Science User Facility supported under Contract DE-AC05-00OR22725. This research was supported by the U.S. Department of Energy, Office of Science, Office of Advanced Scientific Computing Research under Contract No. DE-AC02-05CH11231 by the Accelerated Research for Quantum Computing (I.S. and W.A.d.J.) and Quantum Algorithms Team (M.U.) Programs.
V.P.S. gratefully acknowledges support by the NSF Graduate Research Fellowship Program under Grant No. DGE 1752814, the DOE Office of Science under QuantISED Award DE-SC0019380, and by the BCTP Graduate Student Support Fund.
\bibliographystyle{unsrtnat}
\bibliography{references}
\appendix

\newpage
\section{BKP Model with \texorpdfstring{$h_j=0$}{h\_j=0}}
\label{app:clifford}
One step of time evolution is generated by $U = U_K U_I$ with 
\begin{equation}
   U_{K} = \text{exp}{\left(ib\sum_{j}^{L}\sigma_{j}^{x}\right)} \, , \,  U_{I} = \text{exp}{\left(iJ\sum_{j}^{L-1}\sigma_{j}^{z}\sigma_{j+1}^{z} \right)}.
\end{equation}
In this section, we consider $J = b = \pi/4$ and all the onsite $z$ magnetic fields $h_j$ set to zero.

It is useful to consider the operator dynamics of this model. Take first the action of $U_K$ on single Pauli operators. A short calculation gives
\begin{equation}
    U_K^\dagger X_i U_K = X_i,
\end{equation}
\begin{equation}
    U_K^\dagger Y_i U_K = Z_i,
\end{equation}
and
\begin{equation}
    U_K^\dagger Z_i U_K = - Y_i.
\end{equation}
Note that $U_K$ (and $U_K^{-1}=U_K^\dagger$) maps each Pauli string to another Pauli string, so it is an element of the Clifford group.

For $U_I$, the action on single Paulis is 
\begin{equation}
    U_I^\dagger X_i U_I = - Z_{i-1} X_i Z_{i+1},
\end{equation}
\begin{equation}
    U_I^\dagger Y_i U_I = - Z_{i-1} Y_i Z_{i+1},
\end{equation}
and
\begin{equation}
    U_I^\dagger Z_i U_I = Z_i.
\end{equation}
The action on general Pauli strings follows from these basic formulae. In particular, $U_I$ maps each Pauli string to another Pauli string, so it to is an element of the Clifford group. Hence, $U$ itself is an element of the Clifford group.

Consider also the unitary $Q=e^{i \frac{\pi}{4} Z_L Z_R}$. Note that with $N=3$ qubits on a side and $K=2$ qubits participating in the $V$ interaction, the unitary $e^{i g V}$ with $g = \pi/2$ gives precisely $Q$ for each of the $K$ qubits partipating in $V$. The action of $Q$ on a left Pauli is
\begin{equation}
    Q^\dagger X_L Q = Y_L Z_R,
\end{equation}
\begin{equation}
    Q^\dagger Y_L Q = - X_L Z_R,
\end{equation}
and
\begin{equation}
    Q^\dagger Z_L Q = Z_L.
\end{equation}
Hence, $Q$, and for the right value of $g$ also $e^{i gV}$, is also a element of the Clifford group.

\newpage
\section{Operator Growth}
\label{app:op_growth}
Here we analyze the conditions for successful teleportation in detail by studying the time-dependent operator size. Since we are teleporting a one qubit state, it suffices to consider the operators $\{I,X,Y,Z\}$ acting on the message qubit in the first register on the ``left'' side.
Because the dynamics of the critical BKP model are Clifford when $h_{j} = 0$, the time evolution of Pauli operators results not in a sum of Pauli operators but a single Pauli operator. Below, we write out explicitly the time evolution of these operators. Note that the evolution is also periodic.

\begin{table}[h]

  \centering
  \begin{tabular}{ |c||c|c|c|c|  }
    \hline
    Time & $I_{L}(-t)$ & $X_{L}(-t)$ & $Y_{L}(-t)$ & $Z_{L}(-t)$ \\
    \hline
    t=0   & IIIIII  &  XIIIII &  YIIIII &  ZIIIII    \\
    t=1   & IIIIII  & -ZYIIII &  XYIIII & -YIIIII      \\
    t=2   & IIIIII  & -IZYIII &  XXYIII & -XYIIII      \\
    t=3   & IIIIII  &  IIXIII &  XXZIII & -XXYIII    \\
    t=4   & IIIIII  & -IYZIII &  XZIIII & -XXZIII      \\
    t=5   & IIIIII  & -YZIIII &  ZIIIII & -XZIIII      \\
    t=6   & IIIIII  &  XIIIII & -YIIIII & -ZIIIII      \\
    \hline
  \end{tabular}
\end{table}

Recall that the success criteria for teleportation comes from the phase alignment of operators $\theta_{\mathcal{O}}$. With the form of the coupling as in Eq.~\eqref{eq:interaction}, the phase depends on the $Z$-size of time evolved operators. Whereas the typical size refers to the number of non-identity terms in a Pauli string, by $Z$-size we mean the number of terms that fail to commute with the $Z$ operator. Thus, the $Z$-size measures the number of $X$ and $Y$ terms in a Pauli string.

As a concrete example, at $t=3$, the operator sizes of $\{I, X, Y, Z\}$ are $\{0, 1, 1, 2\}$ respectively. With the value of the two sided coupling of $g=\pi/2$, the phases of $Z$ and $I$ are aligned whereas the phases of $X$ and $Y$ are not. This explains why the effective channel at this special set of parameters is given by teleportation with dephasing noise.

\newpage
\section{WIT circuit compiled for IBM Montreal QPU}
\label{app:compiled-circuit}

\begin{figure}[h]
    \centering
    \includegraphics[scale=0.276]{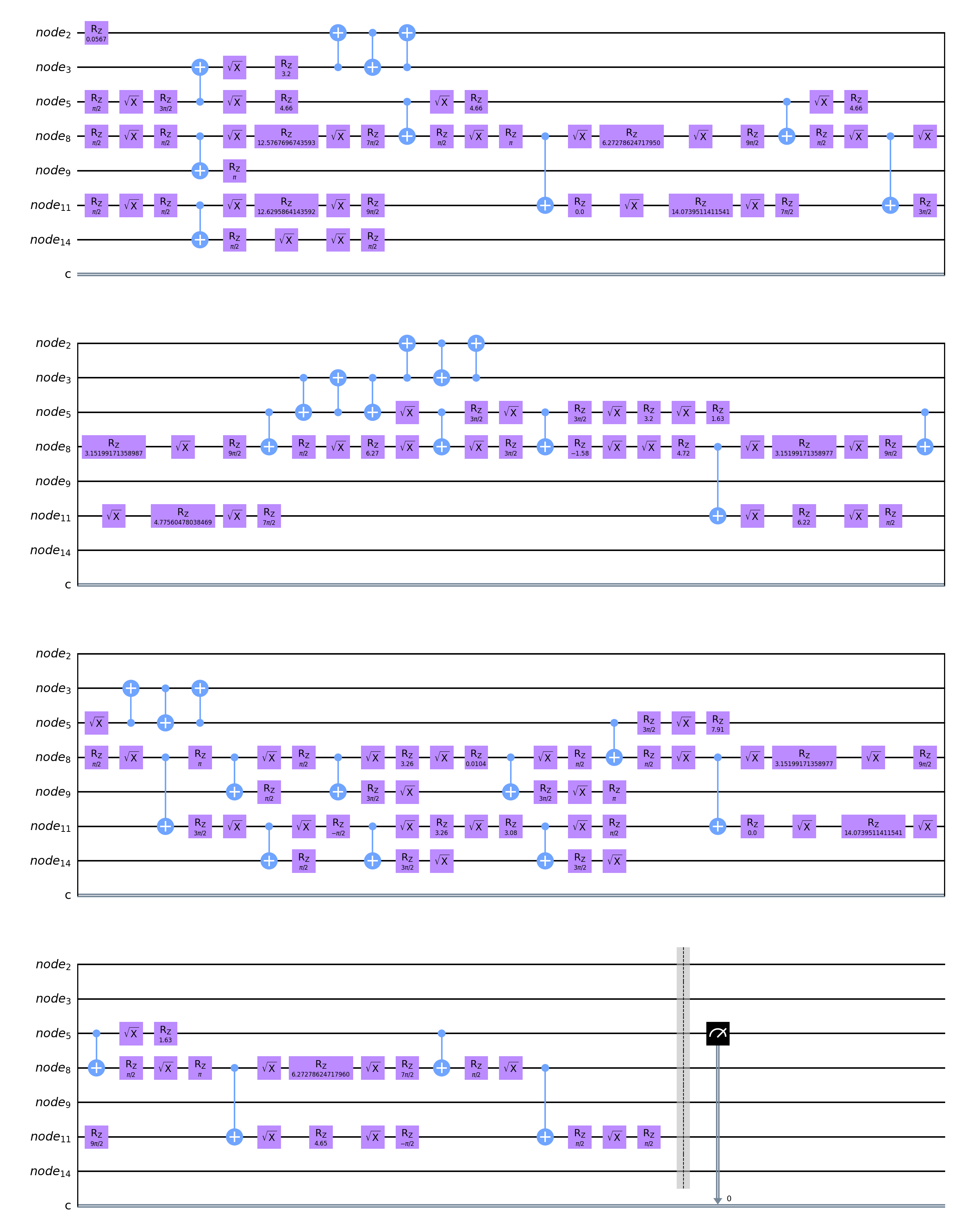}
    \caption{The WIT circuit in Figure~\ref{fig:hl-circuit-8} compiled for \textit{ibmq\_montreal}. Applied compilation passes as per compiler configuration E in Table~\ref{tab:compile-to-IBM}. Nodes are numbered according to the physical map of qubits on the processor. [Drawn with IBM's Qiskit SDK]}
    \label{fig:compiled-circuit-8}
\end{figure}

\newpage
\section{Example of QGLab executing a holographic teleportation experiment}
\label{app:QGLabLog}

\begin{lstlisting}[caption={Log of a typical WIT experiment. This experiment was conducted on \textit{ibmq\_montreal} in advanced compilation mode. Some verbose parts of the log were replaced with ellipsis to make the output more readable.}]

$ wormhole.py --points 14 --retrials 10 \ 
              --qpu ibmq_montreal \
              --toolkit tket \
              --insert-message-with swap \
              --layout-method noise_adaptive \
              --layout select
    
Qubits per side:  3
Time steps:  3
Message insertion method:  swap

Building the library of transpiled circuits...
 ... 1: {q[0]:n[11],q[1]:n[14], ..., q[5]:n[8]}
 ... 2: {q[0]:n[5],q[1]: n[8], ..., q[5]: n[3]}
 ... 3: {q[0]:n[13], q[1]: n[14], ..., q[5]: n[12]}
 ... 4: {q[0]:n[11], q[1]: n[8], ..., q[5]: n[14]}
 ... 5: {q[0]:n[16], q[1]: n[14], ..., q[5]: n[19]}
 ... 6: {q[0]:n[13], q[1]: n[14], ..., q[5]: n[12]}
 ... 7: {q[0]:n[16], q[1]: n[14], ..., q[5]: n[19]}
 ... 8: {q[0]:n[11], q[1]: n[14], ..., q[5]: n[8]}
 ... 9: {q[0]:n[13], q[1]: n[12], ..., q[5]: n[14]}
 ... 10: {q[0]:n[15], q[1]: n[12], ..., q[5]: n[18]}
 
Probing registered initial layouts on default noise model:

 probing layout (1)
 ... running on aer_simulator(ibmq_montreal)
 ... 'run' wall-clock time is 12s
 ... building calibration matrix for readout error 
     mitigation on qubits: [14]
 ... 'get_readout_filter' wall-clock time is 0.01m
 ... building calibration matrix for readout error 
     mitigation on qubits: [11]
 ... 'get_readout_filter' wall-clock time is 0.01m
 probing layout (2)
 ... running on aer_simulator(ibmq_montreal)
 ... 'run' wall-clock time is 12s
 ... building calibration matrix for readout error 
     mitigation on qubits: [9]
 ... 'get_readout_filter' wall-clock time is 0.01m
 ... building calibration matrix for readout error 
     mitigation on qubits: [5]
 ... 'get_readout_filter' wall-clock time is 0.01m
 
 .......
 
 L <Z> med. CX@Meas   CXs/g    Depth/g     Size/g                                   
 8 -0.238  [10] [31,34,34,...] [73,85,...] [157,181,...]   
 9 -0.242  [10] [31,34,34,...] [71,85,...] [156,181,...]   
 7 -0.284  [10] [31,34,34,...] [71,85,...] [156,181,...]   
 0 -0.286  [10] [31,34,34,...] [74,89,...] [160,189,...]   
 4 -0.289  [10] [31,34,34,...] [71,85,...] [156,182,...]   
 5 -0.292  [10] [31,34,34,...] [71,85,...] [156,181,...]   
 3 -0.292  [10] [40,43,43,...] [79,95,...] [159,196,...]   
 6 -0.294  [10] [31,34,34,...] [74,85,...] [159,182,...]   
 2 -0.295  [10] [31,34,34,...] [74,89,...] [156,189,...]   
 1 -0.310  [16] [33,37,37,...] [70,88,...] [165,191,...]
 
 Choose your layout #: 1

Measured qubits across all circuits: [[9], [5]]

Dumping transpiled circuits diagrams
  and QASM representations...

Analyzing transpiled circuits' composition 
  for each point...

 g     Depth Size Gates                                                                  
 0.00  70    165  [('rz', 71), ('sx', 60), ('cx', 33)]   
 0.24  88    191  [('rz', 84), ('sx', 69), ('cx', 37)]   
 0.48  88    187  [('rz', 83), ('sx', 66), ('cx', 37)]   
 0.72  86    189  [('rz', 85), ('sx', 66), ('cx', 37)]   
 0.97  84    185  [('rz', 81), ('sx', 66), ('cx', 37)]   
 1.21  88    190  [('rz', 84), ('sx', 68), ('cx', 37)]   
 1.45  86    191  [('rz', 85), ('sx', 68), ('cx', 37)]   
 1.69  84    183  [('rz', 80), ('sx', 65), ('cx', 37)]   
 1.93  86    186  [('rz', 84), ('sx', 64), ('cx', 37)]   
 2.17  88    193  [('rz', 85), ('sx', 70), ('cx', 37)]   
 2.42  86    192  [('rz', 86), ('sx', 68), ('cx', 37)]   
 2.66  88    190  [('rz', 83), ('sx', 69), ('cx', 37)]   
 2.90  88    190  [('rz', 83), ('sx', 69), ('cx', 37)]   
 3.14  68    169  [('rz', 71), ('sx', 64), ('cx', 33)]   

Sampling Z expectations from quantum experiments 
  on ibmq_montreal...
   ... submitting jobs to ibmq_montreal
Job set name: wormhole_experiment_2021-09-01_16:28:48
          ID: bd6b49...25bb9ff6094143e-16305...0049534
        tags: []
Summary report:
       Total jobs: 1
  Successful jobs: 0
      Failed jobs: 0
   Cancelled jobs: 0
     Running jobs: 0
     Pending jobs: 1

Detail report:
  experiments: 0-139
    job index: 0
    job ID: 61300c...6fa7a68d
    name: wormhole_experiment_2021-09-01_16:28:48_0_
    status: job is being validated

   ... verifying job submissions

   ... waiting for jobs to complete

=> Monitoring 1/1 job:
Job Status: job has successfully run

   ... fetching results
   ... '_fetch_results' wall-clock time is 0.01m
   ... 'run' wall-clock time is 7.31m
   ... building calibration matrix for readout error 
       mitigation on qubits: [9]
   ... 'get_readout_filter' wall-clock time is 0.01m
   ... building calibration matrix for readout error 
       mitigation on qubits: [5]
   ... 'get_readout_filter' wall-clock time is 0.01m

Simulating ideal Z expectations exactly...
   ... 'exact_expectation' wall-clock time is 0.02m

Sampling ideal Z expectations...
   ... running on aer_simulator
   ... 'run' wall-clock time is 0.01m

Sampling Z expectations on mimicked ibmq_montreal...
   ... running on aer_simulator(ibmq_montreal)
   ... 'run' wall-clock time is 0.21m
   ... building calibration matrix for readout error 
       mitigation on qubits: [9]
   ... 'get_readout_filter' wall-clock time is 0.01m
   ... building calibration matrix for readout error 
       mitigation on qubits: [5]
   ... 'get_readout_filter' wall-clock time is 0.01m

Done!

\end{lstlisting}

\end{document}